\documentclass[twocolumn,groupedaddress,aps,nofootinbib,prl,superscriptaddress]{revtex4-1}
\usepackage{amsmath} 
\usepackage{subcaption}
\usepackage{amssymb}
\usepackage{epsfig}
\usepackage{color} 
\usepackage{soul}
 
\begin{document}

\title{$Z_b$ tetraquark channel from lattice QCD and Born-Oppenheimer approximation  } 

\def\UL{Faculty of Mathematics and Physics, University of Ljubljana, Ljubljana, Slovenia}
\def\IJS{Jozef Stefan Institute, Ljubljana, Slovenia}
\def\Regensburg{Institute for Theoretical Physics, University of Regensburg, Regensburg, Germany}
\def\Turkey{Department of Physics, Mimar Sinan Fine Arts University, Bomonti 34380, Istanbul, Turkey}
\author{S. Prelovsek}
 \email{sasa.prelovsek@ijs.si}
   \affiliation{  \UL }
    \affiliation{  \IJS }
\affiliation{\Regensburg}
   \author{H. Bahtiyar }
     \affiliation{  \IJS }
       \affiliation{  \Turkey }
               \author{J. Petkovi\'  c }
      \affiliation{     \UL } 
       \affiliation{     \IJS } 

\begin{abstract} 

 Two $Z_b$ hadrons with exotic quark structure $\bar bb\bar du$ were discovered by Belle experiment. We present  a lattice QCD study of the $\bar bb\bar du$ system  in the approximation of static $b$ quarks, where the total spin of heavy quarks is fixed to one.    The energies of eigenstates   are determined as a function of  the separation $r$ between $b$ and $\bar b$.  The lower eigenstates  are related to a bottomonium and a pion.
 The eigenstate dominated by $B\bar B^*$ has energy  significantly below $m_B+m_{B^*}$, which points to  a sizable attraction for small $r$.  The  attractive potential $V(r)$ between $B$ and $\bar B^*$ is extracted assuming  that this eigenstate is related exclusively to $B\bar B^*$. The  Schr\"odinger equation for   $B\bar B^*$ within the extracted potential leads to one  bound state below $B\bar B^*$ threshold, whose mass depends on the  parametrization of the lattice potential. For certain parametrizations, 
 the bound state is very close to the $B\bar B^*$ threshold and renders  a narrow peak in the $B\bar B^*$ rate above threshold  - these features could be  related to $Z_b(10610)$ in the experiment.       \end{abstract}
  
\maketitle

The Belle experiment discovered two  $Z_b^+$ states with exotic quark content $\bar bb\bar du$,  $J^P\!\!=\!\!1^+$  and $I\!\!=\!\!1$ in 2011 \cite{Belle:2011aa,Garmash:2014dhx}.  The lighter  $Z_b(10610)$ lies slightly above $B\bar B^*$ threshold and the heavier $Z_b(10650)$ just above $B^*\bar B^*$. The observed  decay modes  are
$\Upsilon(1S,2S,3S)\pi^+,\  h_b(1P,2P)\pi^+,\ B\bar B^*$ and $B^*\bar B^*$
  \cite{Belle:2011aa,Garmash:2014dhx,Garmash:2015rfd}, 
 where the $B\bar B^*$ and  $B^*\bar B^*$ largely dominate   $Z_b(10610)$ and $Z_b(10650)$ decays, respectively.   Many phenomenological theoretical studies of these two states have been performed, for example \cite{Wang:2018jlv,Kang:2016ezb,Ortega:2019uuk,Wang:2018pwi,Yang:2017rmm,Voloshin:2017gnc,Goerke:2017svb,Dias:2014pva,Guo:2016bjq,Ali:2011ug,He:2014nya,Karliner:2013dqa,Cleven:2013sq,Esposito:2016itg}, and the majority indicates that $B^{(*)}\bar B^*$ Fock component   is important. 
 
We explore this channel within the first-principle  lattice QCD.  The only preliminary lattice  study of this channel has been reported in \cite{Peters:2016wjm,Peters:2017hon} and is reviewed below.   No other lattice results  are available since this channel presents a severe challenge. 
Scattering matrix would have to be determined using the L\"uscher method for  at least 7 coupled two-meson channels listed in the previous paragraph. Poles of the scattering matrix would render possible $Z_b$ states. Following this path seems too challenging at present. Furthermore, the original L\"uscher approach for two-particle scattering is not valid above the three-particle threshold.  
 
 In the present study we consider the  Born-Oppenheimer approximation \cite{BO}, inspired by the study of this system in \cite{Peters:2016wjm,Peters:2017hon}.    It is   applied in molecular physics since ions are much heavier than other degrees of freedom.   It is valuable also for the $Z_b$ system  $\bar bb\bar du$, where $b$ and $\bar b$ represent heavy degrees of freedom ($h$), while the light quarks and gluons are light degrees of freedom ($l$), see for example \cite{Braaten:2014qka,Brambilla:2017uyf}. The simplification comes from the fact that the heavy degrees of freedom have large mass and therefore small velocity and kinetic energy.  
 
\begin{figure*}[ht!]
\begin{subfigure}{0.2\textwidth}
\includegraphics[width=\linewidth]{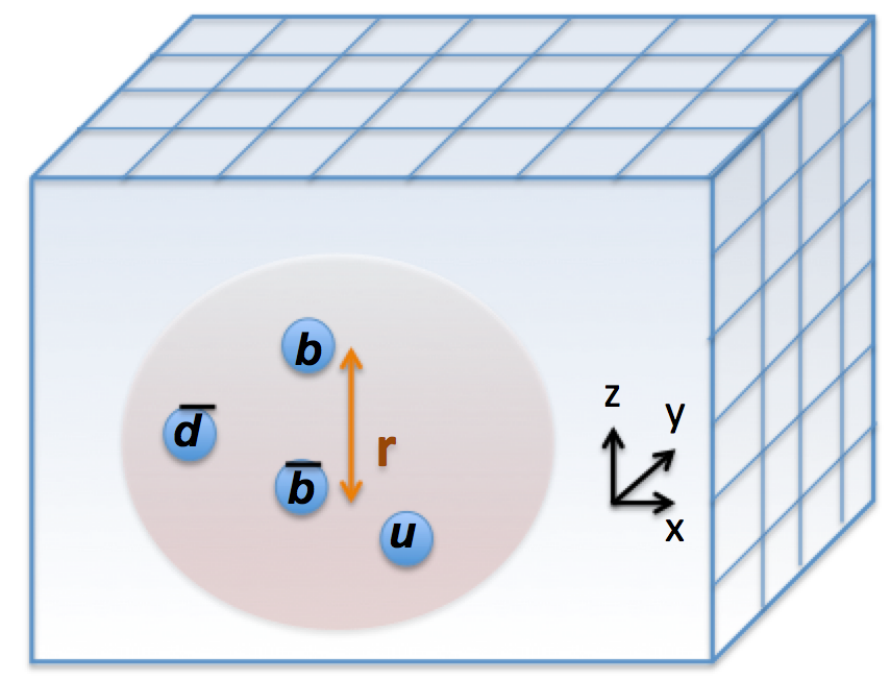}
\caption{ } \label{fig:1a}
\end{subfigure}
\hspace*{\fill} 
\begin{subfigure}{0.2\textwidth}
\includegraphics[width=\linewidth]{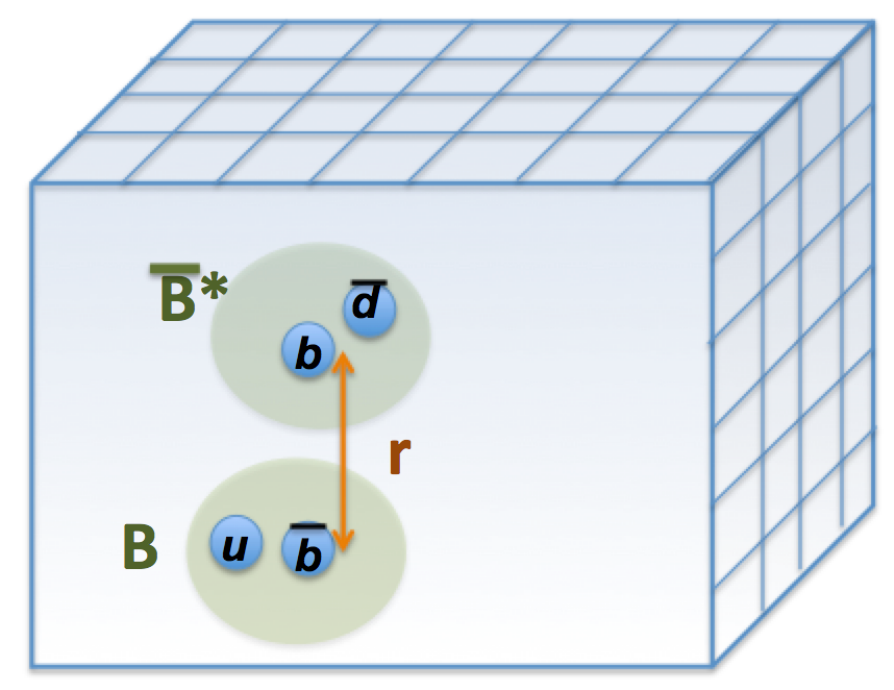}
\caption{ } \label{fig:1b}
\end{subfigure}
\hspace*{\fill} 
\begin{subfigure}{0.2\textwidth}
\includegraphics[width=\linewidth]{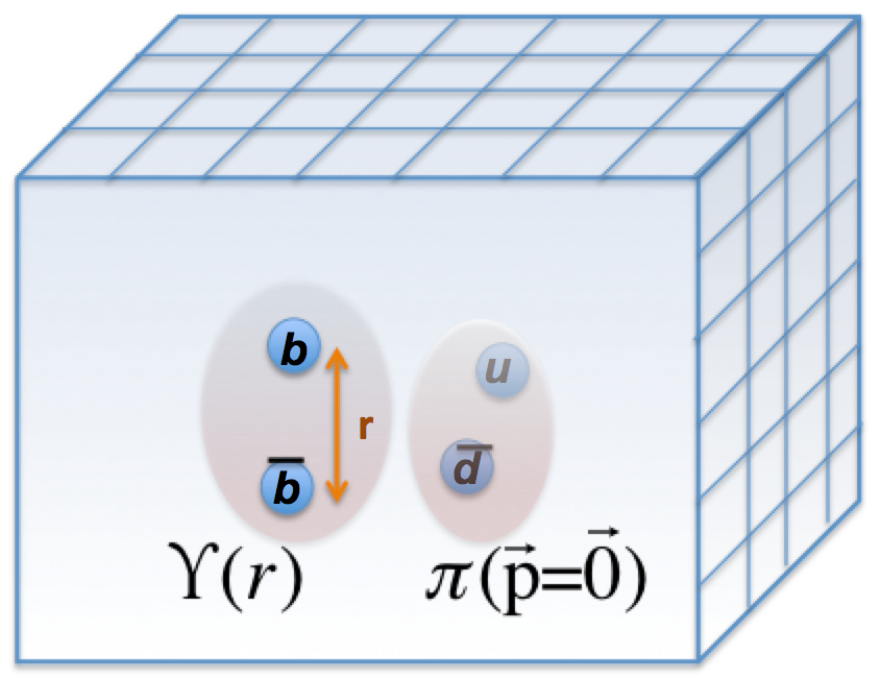}
\caption{ } \label{fig:1c}
\end{subfigure}
\hspace*{\fill} 
\begin{subfigure}{0.2\textwidth}
\includegraphics[width=\linewidth]{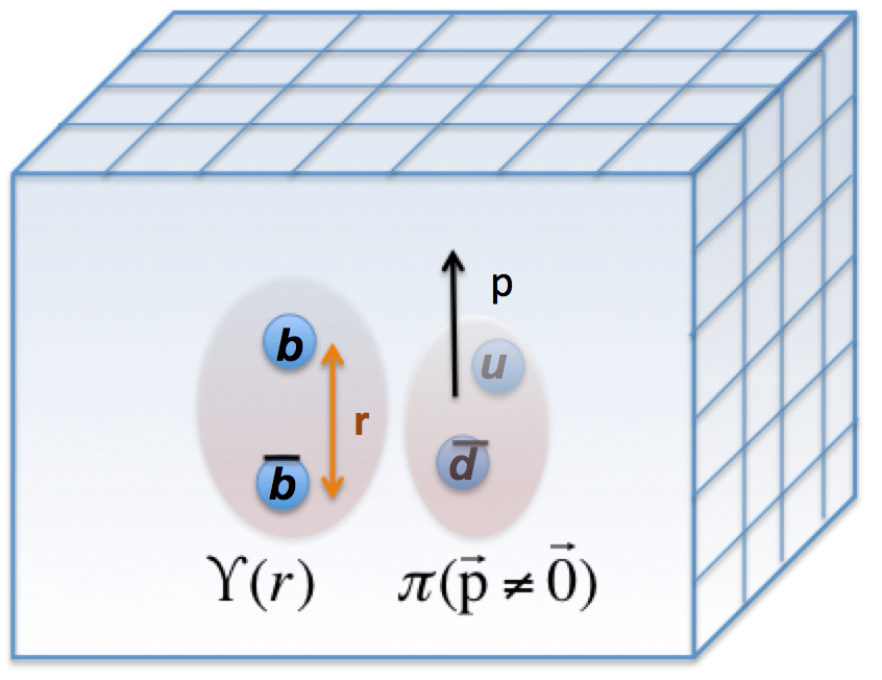}
\caption{ } \label{fig:1d}
\end{subfigure}
\caption{ (a) The system considered. (b-d) Two-hadron Fock components in the system with quantum numbers (\ref{sym}). } \label{fig:1}
\end{figure*}

  In the first step we treat $b$ and $\bar b$ as static  at fixed   distance $r$ (Figure \ref{fig:1a}) and the main purpose is to  determine    eigen-energies  $E_n(r)$ of this system.   This energy   represents the total energy without the kinetic and rest energies  of the $b$ and $\bar b$, so $E_n(r)$ is   related to the potential  $V(r)$ felt by the heavy degrees of freedom. In the second step,   we study the motion of the heavy degrees of freedom (with the physical masses)  under the influence of the extracted potential $V(r)$.  Solutions of the  Schr\"odinger equation  render information on possible  (virtual) bound states $Z_b$,  resonances and cross-sections. 
 
  The low-lying eigenstates  of the system in Fig. \ref{fig:1a}   with quantum numbers  (\ref{sym}) are related to two-hadron states in Figs. \ref{fig:1} (b-d)
 \begin{equation}
 \label{two-hadron-states}
 B(0)\bar B^*(r), ~  \Upsilon(r)\pi(\vec p=0),  ~ \Upsilon(r)\pi(\vec p\not =0),  ~ \Upsilon(r)b_1(\vec 0).   
 \end{equation} 
   The eigen-energy $E_n(r)$ related to $B\bar B^*$   in Fig.  \ref{fig:1}b   is  of major interest  since $Z_b$   lies near $B\bar B^*$ threshold \cite{Belle:2011aa}.     The $\Upsilon(r)\pi(\vec p)$ represent the ground state at small $r$. Here $\Upsilon(r)$ denotes the spin-one bottomonium where $\bar b$ and $b$ are separated by $r$.  Pion can have zero or non-zero momentum  $\vec p=\vec n \tfrac{2\pi}{L}$ since the total momenta of light degrees of freedom is not conserved in the presence of static quarks, i.e. pion momentum can change   when it scatters on an infinitely heavy  $\Upsilon$. 
     Our task is to extract energies of all these eigenstates $E_n(r)$ as a function of   $r$. 
   The only previous lattice study of this system  \cite{Peters:2016wjm} presents preliminary results based on  Fock components  $B\bar B^*$ and $\Upsilon \pi(0)$; the presence   $\Upsilon \pi(\vec p\not =0)$  was mentioned in  \cite{Peters:2017hon}, but not included in the simulation.

\vspace{0.1cm}

   {\it Quantum numbers and operators:}
   We consider $Z_b^0$ that has   quantum numbers $I\!=\!1$, $I_3\!=\!0$, $J^{PC}\!=\!1^{+-}$ and $J_z\!=\!0$ in experiment.   The list of conserved quantum numbers is slightly different in the systems with two static particles. We study the system in Fig. \ref{fig:1a}  with quantum numbers  
      \begin{align}
   \label{sym}
&I=1,\ I_3=0, \   \epsilon=-1,\  C\cdot P=-1~\\
&S^{h}=1,\ S_z^{h}=0,\ J_z^{l}=0,\ (h=\mathrm{heavy},~ l=\mathrm{light})\nonumber
   \end{align}
   where the neutral system is considered where C-conjugation can be applied (Fig. \ref{fig:1} shows the charged partner).  Only the z-component of angular momenta for the light degrees of freedom ($J_z^{light}$) is conserved. The quantum number  $\epsilon$  corresponds to the reflection over the yz plane. $P$ refers to inversion with respect to mid-point between $b$ and $\bar b$ and $C$ is charge conjugation, where only their product is conserved. The quantum numbers    in  (\ref{sym}) are  conventionally  denoted  by $\Sigma^-_u$ using the conventions from \cite{Juge:1999ie}\footnote{ This  provides irreducible representation  $(J_z^l)^\epsilon_{CP}=\Sigma^-_u$, where the  notation here and in   \cite{Juge:1999ie} is related by  $J_z^l\to \Lambda$, $J_z^l=0\to \Sigma$, $CP\to \eta$, $CP=-1\to u$ .}.
   
   The spin  of the infinitely heavy quark   can not flip via the interaction with gluons, so spin $S^{h}$ of $\bar bb$  is conserved. We choose to study  the system with $S^{h}\!=\!1$, which can decay to $\Upsilon$, while it can not decay to  $\eta_b$ and $h_b$  since these carry $S^{h}\!=\!0$. Note that the physical   $Z_b$ and $B\bar B^*$ with finite $m_b$ can be a linear combination of $S^{h}\!=\!1$ as well as $S^{h}\!=\!0$,   and we  study only $S^{h}\!=\!1$ component here. We have in mind this  component, which includes $B\bar B^*$, $\bar BB^*$, $\bar B^* B^*$ ($O_1$ in Eq. \ref{operators}), when we refer to "$\!B\bar B^*$" throughout this paper. 
   
      The eigen-energies $E_n$ of the system   in Fig. \ref{fig:1a} are determined from the correlation functions $\langle O_i (t)O_j^\dagger(0)\rangle$.  We employ 6 operators $O_i$ that create/annihilate the system with quantum numbers (\ref{sym}) and resemble Fock components (\ref{two-hadron-states})  in Figs. \ref{fig:1} (b-d)
    \begin{align}
   \label{operators}
&O_1\!=\! O^{B\bar B^*}\!\!\propto\sum_{a,b}\sum_{A,B,C,D}\!\!\!\!\Gamma_{BA}\tilde \Gamma_{CD}~\bar b^a_C(0)q_A^a(0)~ \bar q^b_B(r)b_D^b(r) \nonumber \\ 
&\propto \bigl([\bar b(0) P_- \gamma_5 q(0)]~[\bar q(r) \gamma_z P_+ b(r)]+\{\gamma_5\leftrightarrow \gamma_z\}\bigr)   \nonumber \\ 
& \ + \bigl( [\bar b(0) P_- \gamma_y q(0)]~[\bar q(r) \gamma_x P_+ b(r)]- \{\gamma_y\leftrightarrow \gamma_x\}\bigr)  \nonumber \\
&O_2\!=\! O^{B\bar B^*} \nonumber\\
&O_3\!=\!O^{\Upsilon \pi(0)}\!\propto\! [\bar b(0) U\gamma_z P_+ b(r)]~[\bar q \gamma_5q]_{\vec p=\vec 0}\nonumber\\ &O_4\!=\!O^{\Upsilon \pi(1)}\!\propto\! [\bar b(0) U\gamma_z P_+ b(r)]~ \bigl([\bar q \gamma_5q]_{\vec p=\vec e_z}+ [\bar q \gamma_5q]_{\vec p=-\vec e_z}\bigr)\nonumber\\ 
&O_5\!=\!O^{\Upsilon \pi(2)}\!\propto\! [\bar b(0) U\gamma_z P_+ b(r)]~ \bigl([\bar q \gamma_5q]_{\vec p=2\vec e_z}+ [\bar q \gamma_5q]_{\vec p=-2\vec e_z}\bigr)\nonumber\\  
&O_6\!=\!O^{\Upsilon b_1(0)}\!\propto\! [\bar b(0) U\gamma_z P_+ b(r)]~[\bar q \gamma_x\gamma_y q]_{\vec p=\vec 0} ~.
\end{align}
Here $\Gamma\!\!=\!\!P_-\gamma_5$, $\tilde \Gamma\!=\!\gamma_z P_+$, $[\bar q \Gamma^\prime q]_{\vec p}\equiv \tfrac{1}{V} \sum_{\vec x} \bar q(\vec x) \Gamma^\prime q(\vec x) e^{i \vec p\vec x}$, momenta is given in units of $2\pi/L$, capital (small) letters represent Dirac (color) indices, color singlets are denoted by $[..]$ and $U$ is a product of gauge links between $0$ and $r$. 
First line in $O_{1}$  decouples spin indices of light and heavy quarks in order to make $J^{l}_z$ and $S^{h}_{(z)}$ (\ref{sym}) are more transparent  \cite{Peters:2016wjm}, while the second line  is obtained via the Fierz transformation. $O_2$ is obtained from $O_1$ by replacing all $q(x)$ with $\nabla^2 q(x)$. $O_{4,5}$ have pion momenta  in $z$ direction due to $J^{l}_z=0$ and have two terms to ensure $C\cdot P=-1$.   The $\Upsilon b_1$ is not a decay mode  for finite  $m_b$ where $C$ and $P$ are separately conserved, but it is has quantum numbers (\ref{sym}) for $m_b\to \infty$. 
The pair $\bar qq$  indicates combination $\bar uu-\bar dd$ with $I=1$ and $I_3=0$.  
 All light quarks $q(x)$ are smeared around the position $x$ using the full distillation \cite{Peardon:2009gh} with  the radius about $0.3~$fm, while the heavy quarks are point-like. 

We verified there are no other two-hadron states in addition  to (\ref{two-hadron-states}) with quantum numbers (\ref{sym}) and with non-interacting energies  (\ref{Eni}) below $m_B+m_{B^*}+0.2~$GeV.  

    \begin{figure}[h!]
	\begin{center}
	\includegraphics[width=0.47\textwidth]{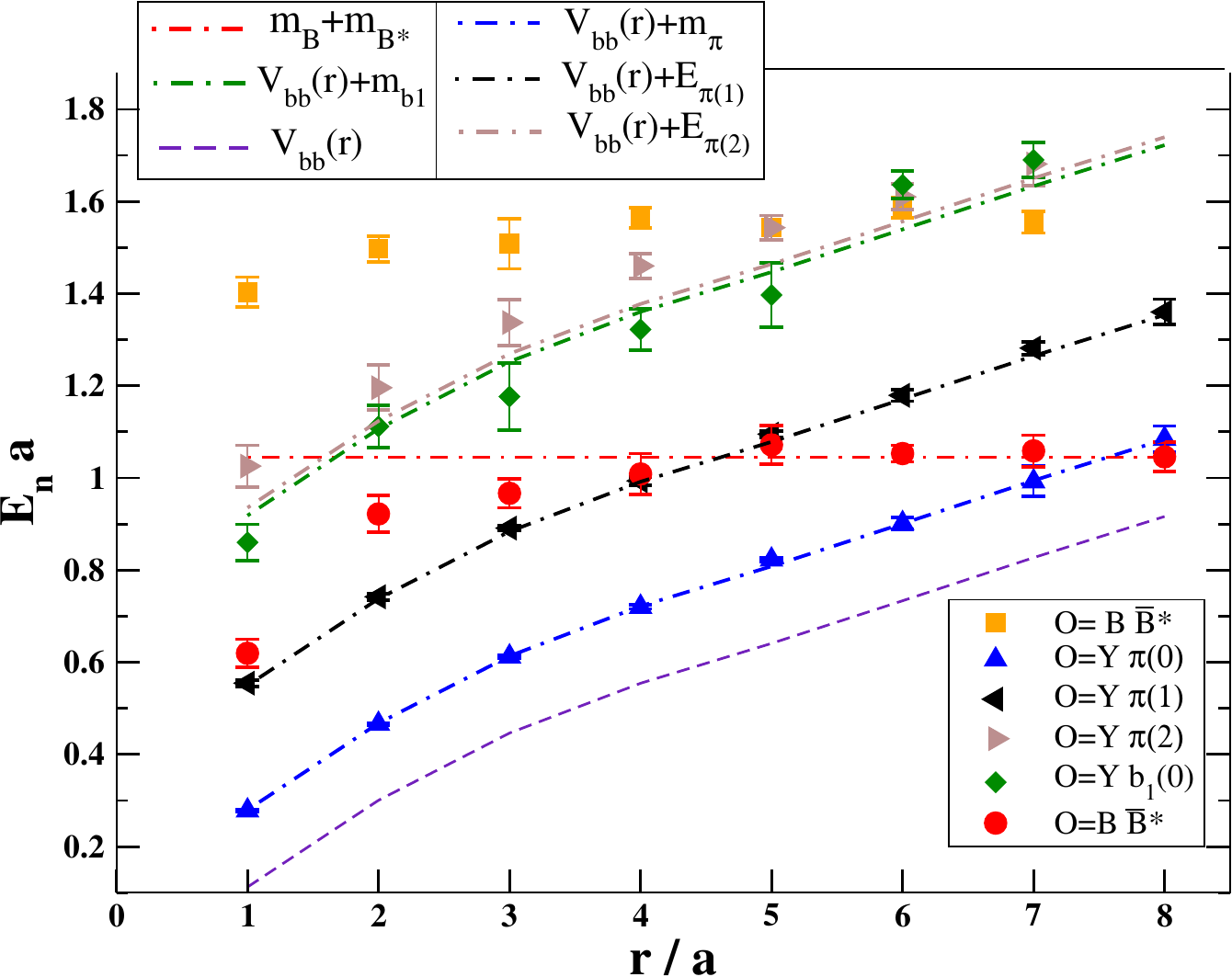}   
		\caption{ Eigen-energies of  $\bar bb\bar du$  system (Fig. 1a) for various separations $r$ between  static quarks $b$ and $\bar b$ are shown by points. The  label  indicates which two-hadron component dominates each eigenstate. The dot-dashed lines represent related two-hadron energies $E^{n.i.}$ (\ref{Eni}) when two hadrons  (\ref{two-hadron-states}) do not interact. The eigenstate dominated by $B\bar B^*$ (red circles) has energy  significantly below $m_B+m_{B^*}$ and  shows  sizable  attraction. Lattice spacing is $a\simeq 0.124~$fm.  }
	\label{fig:En}
	\end{center}
\end{figure}

 \vspace{0.1cm} 
  
  {\it Lattice details:}  Simulation is performed on   an ensemble with dynamical  Wilson-clover $u/d$ quarks,  $m_\pi \simeq 266(5)~$MeV, $a\simeq 0.1239(13)~$fm and 280 configurations \cite{Hasenfratz:2008ce,Lang:2011mn}.     We choose an ensemble with small $N_L\!=\!16$ and $L\!\simeq\! 2~$fm so that  $\Upsilon \pi(p_z)$ with   $p_z> 2 \tfrac{2\pi}{L}$ appear at  $E\!>\! m_B+m_{B^*}+0.2~$GeV above our interest; larger $L$ would require  further operators like $O_{4,5}$ with higher $\vec p$. Small $L$ 
 restricts us to   $r/a\leq \tfrac{1}{2}N_L= 8$,  but the statistical errors grow with  $r$ and the current precision prevents us from accurate  results for $r/a>8$ anyway.   Small $L$ leads also to the usual exponentially-suppressed corrections related to the pion and a very mild effect on the light-quark cloud in a $B$-meson since $r_B\ll L$.  The lattice temporal extent $N_T\!=\!32$ is effectively  doubled by summing the light-quark propagators with periodic and anti-periodic boundary conditions in time \cite{Lang:2011mn}. 
 
  \vspace{0.1cm} 

{\it Calculation of eigen-energies  and overlaps:}  Correlation matrices $C_{ij}(t)=\langle O_i (t)O_j^\dagger(0)\rangle$ are evaluated using the full distillation method  \cite{Peardon:2009gh}.  The $\bar bb$ annihilation Wick contraction is not present in the static limit considered here.    $C_{ij}$ are averaged over $8^3$ or $16^3$ space positions of $\bar b$, while sub-matrix for $O_{3-6}$ is averaged over all source time slices to increase accuracy.   Eigen-energies $E_n$ and overlaps $\langle O_i|n\rangle$ are extracted from  the $6\times 6$ matrices $ C_{ij}(t)=\sum_n \langle O_i|n\rangle e^{-E_nt} \langle n|O_j^\dagger\rangle$ using the widely used GEVP variational  approach \cite{Michael:1985ne,Luscher:1990ck,Blossier:2009kd}.   

 \vspace{0.2cm}
 
 {\it Eigen-energies of  $\bar bb\bar du$  system as a function of $r$:}  
   The main result of our study are the  eigen-energies of the $\bar bb\bar du$  system (Fig.  \ref{fig:1}a) with static $b$ and $\bar b$ separated by $r$. They are shown by points in Figure \ref{fig:En}.  The colors of points indicate which Fock-component  (\ref{two-hadron-states}) dominates an eigenstate, as determined from the normalized overlaps  of an eigenstate $|n\rangle$ to operators $O_i$.  Normalized overlap $\tilde Z_i^n\equiv \langle O_i|n\rangle/\max_m \langle O_i|m\rangle$ is normalized so that its maximal value for given $O_i$  across all eigenstates is equal to one.

   The dashed lines in Fig. \ref{fig:En} provide the related non-interacting (n.i.) energies $E_n$ of two-hadron states (\ref{two-hadron-states}) 
   \begin{equation}
   \label{Eni}
  E^{n.i.}_{B\bar B^*}\!=\!2m_B,\ 
  E^{n.i.}_{\Upsilon\pi(\vec p)}\!=\!V_{\bar bb}(r)+E_{\pi(\vec p)},\  
  E^{n.i.}_{\Upsilon b_1(0)}\!=\!V_{\bar bb}(r)+m_{b_1},\    
  \end{equation} 
   where $\bar bb$ static potential $V_{\bar bb}(r)$,  $E_{\pi(\vec p)}\simeq \sqrt{m_\pi^2+\vec p^2}$, $m_{b_1}$ and $m_B=m_{B^*}=0.5224(14)$ (mass of $B^{(*)}$ for $m_b\to \infty$ without $b$ rest mass) are determined on the same lattice.  
          
      The eigenstate dominated by $B\bar B^*$ has an energy close to $m_B+m_{B^*}$ for $r>0.5~$fm, but it has significantly lower energy for $r\simeq [0.1,0.4]~$fm (red circles in Fig. \ref{fig:En}). This indicates sizable strong attraction between $B$ and $\bar B^*$ in this system - something that might   be related to the existence of $Z_b$ tetraquarks. This is the most important and robust result of this lattice study.

Other eigenstates are dominated by $\Upsilon \pi(\vec p)$ and $\Upsilon b_1$. Their  energies $E$  lie close to the non-interacting energies $E^{n.i.}$ (\ref{Eni}) given by dot-dashed lines, so $E\simeq E^{n.i.}$. We point out that we can not claim nonzero energy shifts $E-E^{n.i.}$ for $\Upsilon \pi$ and $\Upsilon b_1$ states (although Fig.   \ref{fig:En} shows small deviations 
from zero in some cases) since the statistical and systematic errors  are not small enough.   

\vspace{0.1cm}

{\it Towards masses of $Z_b$ states within certain approximations:} Eigen-energies of $\bar bb\bar du$ system in Fig. \ref{fig:1a} indicate that eigenstate dominated by the $B\bar B^*$  has significantly lower energy than $m_B+m_{B^*}$ at small separation $r$ between static $b$ and $\bar b$. This suggests a possible existence of exotic hadron (related to $Z_b$)  and  related peak in the cross-section  near $B\bar B^*$ threshold. Such physical observables require the study of the motion for the heavy degrees of freedom based on the energies $E_n(r)$ according to the Born-Oppenheimer approach.   The  precise prediction of  such observables is not possible at present since lattice eigen-energies  are not known for $r\!<\!a$. In addition, the accurate study   would require the coupled-channel treatment of  all Fock components (\ref{two-hadron-states})  through the coupled-channel Schr\"odinger equation, which is a challenging task left for the future  (this  was recently elaborated in \cite{Bicudo:2019ymo} for conventional  $\bar bb$ with $I\!=\!0$). 
 
We apply two simplifying approximations in order to shed light on the possible existence of $Z_b$ based on energies  in Figure \ref{fig:En}. The first assumption is  that the eigenstate indicated by red circles in Fig. \ref{fig:En} is related exclusively to $B\bar B^*$ Fock component and does not contain other Fock components in (\ref{two-hadron-states}). This is supported  by our lattice results
 to a very good approximation, since this eigenstate couples almost exclusively to  $O^{B\bar B^*}$  and has much smaller coupling to $O^{\Upsilon \pi}$ and $O^{\Upsilon b_1}$: the normalized overlap    of this state to  $O^{\Upsilon \pi,\Upsilon b_1}$ is $\tilde Z_{3-6}\leq0.07$ for $r/a\leq 3$, while overlap to $O^{B\bar B^*}$ is $\tilde Z_{1,2}\simeq {\cal O}(1)$. In the reminder we explore the  physics implications of this eigen-energy $E_{B\bar B^*}(r)$.

\begin{figure}[tb!]
\begin{center} 
\includegraphics[width=0.49\textwidth]{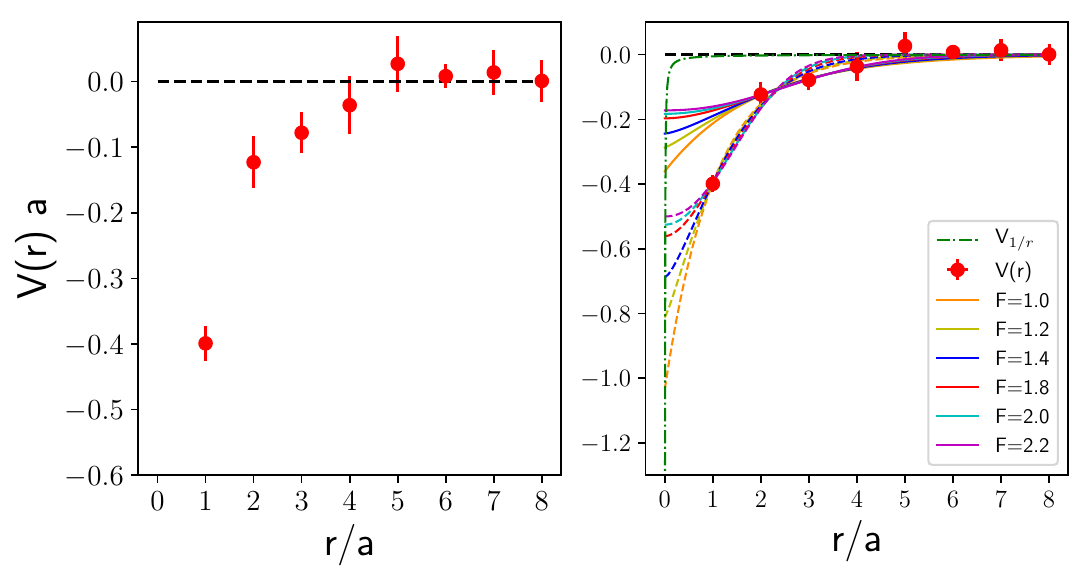}  
\caption{ (a) The   extracted potential $V(r)$ between $B$ and $\bar B^*$ from lattice. (b) Fits  of  $V(r)$ assuming the form of the regular potential $V_{reg}$ (\ref{V}) are presented by the solid and dashed lines  for various values of parameter $F$. Solid and dashed lines correspond to fits in the ranges $r/a=[2,4]$ and $[1,4]$, respectively.    
The singular potential $V_{1/r}(r)$ is shown by dot-dashed green line.  Lattice spacing is $a\simeq 0.124~$fm.   } \label{fig:V}
\end{center}
\end{figure}

 The energy $E_{B\bar B^*}(r)$ represents   the total energy without  the kinetic energy of heavy degrees of freedom. The difference $V(r)=E_{B\bar B^*}(r)-m_B-m_{B^*}$ therefore represents the potential felt by the heavy degrees of freedom, in this case between $B$ and $\bar B^*$ mesons. The extracted potential is plotted in Fig. \ref{fig:V}. 
  The potential shows sizable attraction for $r=[0.1,0.4]~$fm and is compatible with zero for $r\geq 0.6~$fm  within sizable errors.  Lattice study that would probe whether one-pion exchange dominates at large $r$ would    need higher accuracy. 
 
The problem is that the potential $V(r)$ is not determined from the lattice for $r\!<\!a$, it might be affected by discretization effects at $r\!\simeq\!  a$ and  the analytic form of r-dependence is not known apriori. This brings us to the second simplifying approximation
 \begin{equation}
 \label{V}
 V(r)=V_{reg.}(r)+V_{1/r}(r),\   V_{reg.}(r)=  -A\, e^{-(r/d)^F} ,
  \end{equation}
  where we assume a certain form of the regular potential  $V_{reg.}(r)$  that has no singularity at $r\to 0$. The fits of the lattice potential for   various choices of the parameter $F$  (\ref{V})  are shown in Fig. \ref{fig:V}. We employ two choices of fitting ranges   $r/a=[2,4]$ or $[1,4]$  since the  lattice potential at $r/a=1$ can be prone to the lattice discretization errors.    The question if the potential contains also a singular piece $1/r$ can be addressed perturbatively, giving   $V_{1/r}^{{\cal O}(\alpha_s)}(r)=0$ and  $V_{1/r}(r)=\tfrac{1}{9}[V_0(r)+8V_8(r)]=\tfrac{ \delta a_2}{108 \pi^2 }    \tfrac{\alpha_s^3}{r}  $ \cite{Kniehl:2004rk} for very small $r$. This follows from the interaction of $\bar b$ and $b$ within $B\bar B^*$, while other pairs among $\bar bb\bar qq$ are at average   distance of the order of B-meson size and do not lead to singularity at $r\to 0$.  Results below are based on $V_{reg}+V_{1/r}$; we have verified that masses and cross-sections based solely on $V_{reg}$ agree within the errors since $V_{1/r}$ is suppressed.  
  
The motion of $B$ and $\bar B^*$ within the extracted   potential $V(r)$ is analyzed by solving the non-relativistic 3D Schr\"odinger equation $[-\frac{1}{2\mu}\tfrac{d^2}{dr^2}+\tfrac{l(l+1)}{2\mu r^2}+V(r)]u(r)=Wu(r)$  for the experimentally measured  $B^{(*)}$ meson masses and $1/\mu=1/m_B^{exp}+1/m_{B^*}^{exp}$.    Here  $W=E^{tot}-m_B-m_{B^*}$ is the energy with respect to $B\bar B^*$ threshold. The $B$ and $\bar B^*$ can couple to $Z_b$ channel  with $J^P\!=\!1^+$ in partial waves $l=0,2$.  Below we extract (virtual) bound  states  and scattering rates for  $l=0$, while $l=2$ is not discussed  since   $V(r)+\tfrac{l(l+1)}{2\mu r^2}>0$   is repulsive for all $r$. 

The wave functions of the Schr\"odnger equation render the phase shift $\delta_{l=0}(W)$ and $B\bar B^*$ scattering matrix $S(W)=e^{2i\delta_{0}(W)}$. Resonances above threshold do not occur for purely attractive s-wave potentials since there is no barrier to keep the state metastable, while (virtual) bound states below threshold may be present. 
Bound state (virtual bound state) corresponds to the pole of  $S(W)$ for real  $W<0$ and imaginary momenta $k=i|k|$ ($k=-i|k|$) of $B$   in the center of momentum frame.  

 \begin{figure}[tb!]
\begin{center} 
\includegraphics[width=0.5\textwidth]{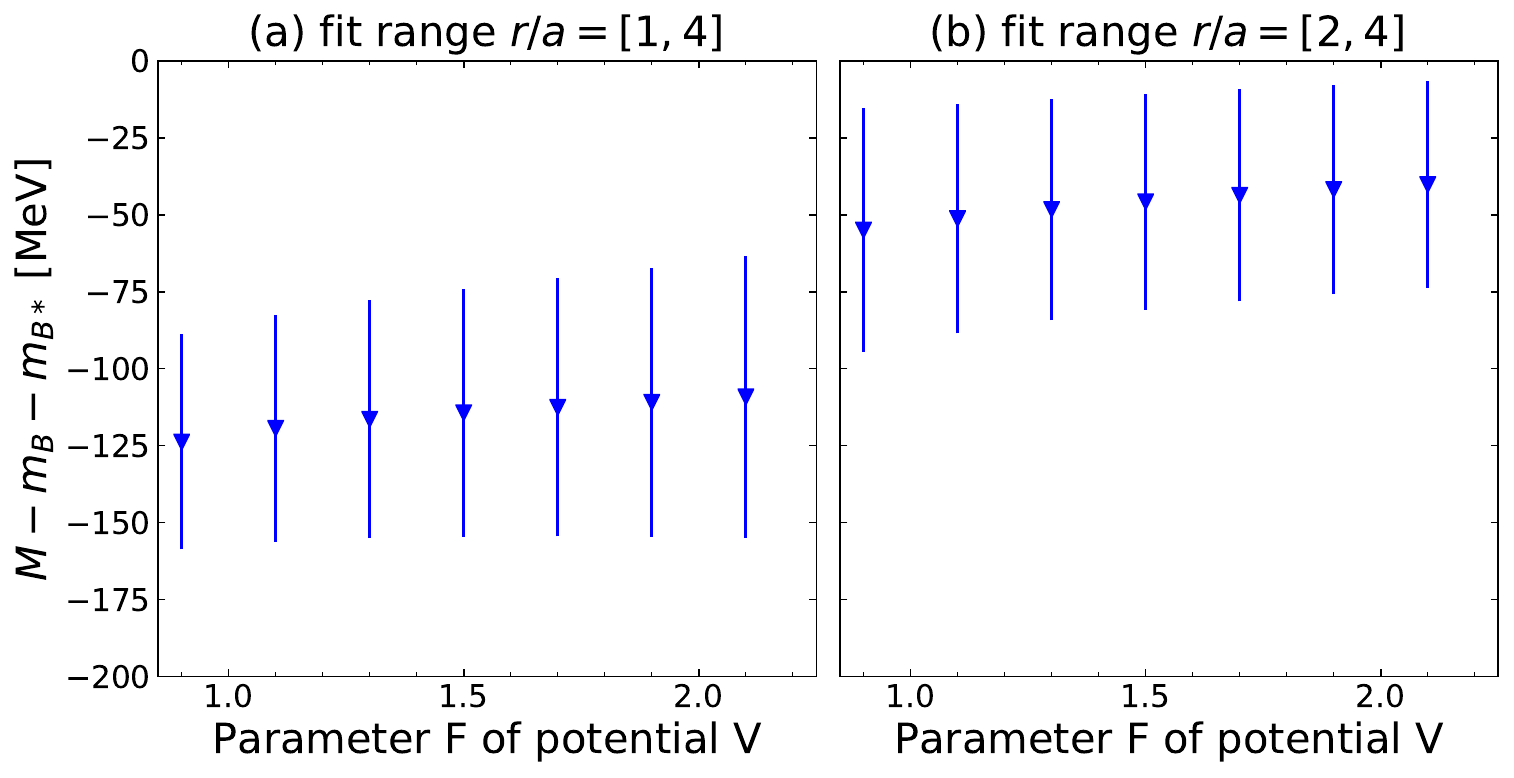}  
\caption{ The mass of the  bound state for various fits of the potential $V(r)$.   The mass is shown  for various choices of the parameter $F$ in $V(r)$ (\ref{V})  and the fitting range in $r$. } \label{fig:masses}
\end{center}
\end{figure}

We find one bound state   below threshold and its mass is shown   in Fig.    \ref{fig:masses}.  
The mass is presented for various choices of  the parameter $F$ in  the potential  (\ref{V}) and the fitting ranges in $r$. 
 The bound state with mass $M$ lies at  
\begin{equation}
\label{mass}
M-m_B-m_{B^*}=- 48^{~+41}_{~-108} ~\mathrm{MeV} ~. 
\end{equation}
 The central value corresponds to  fit of the potential (\ref{V}) in the range $r/a=[2,4]$ that renders the parameters $F=1.3$, $A=0.262(38)$, $d=2.51(27)$. The significant uncertainty of the binding energy captures the statistical errors  
 as well as  various choices for  parametrizing the potential in Fig.  \ref{fig:masses}. 
 
\begin{figure} 
\includegraphics[width=0.4\textwidth]{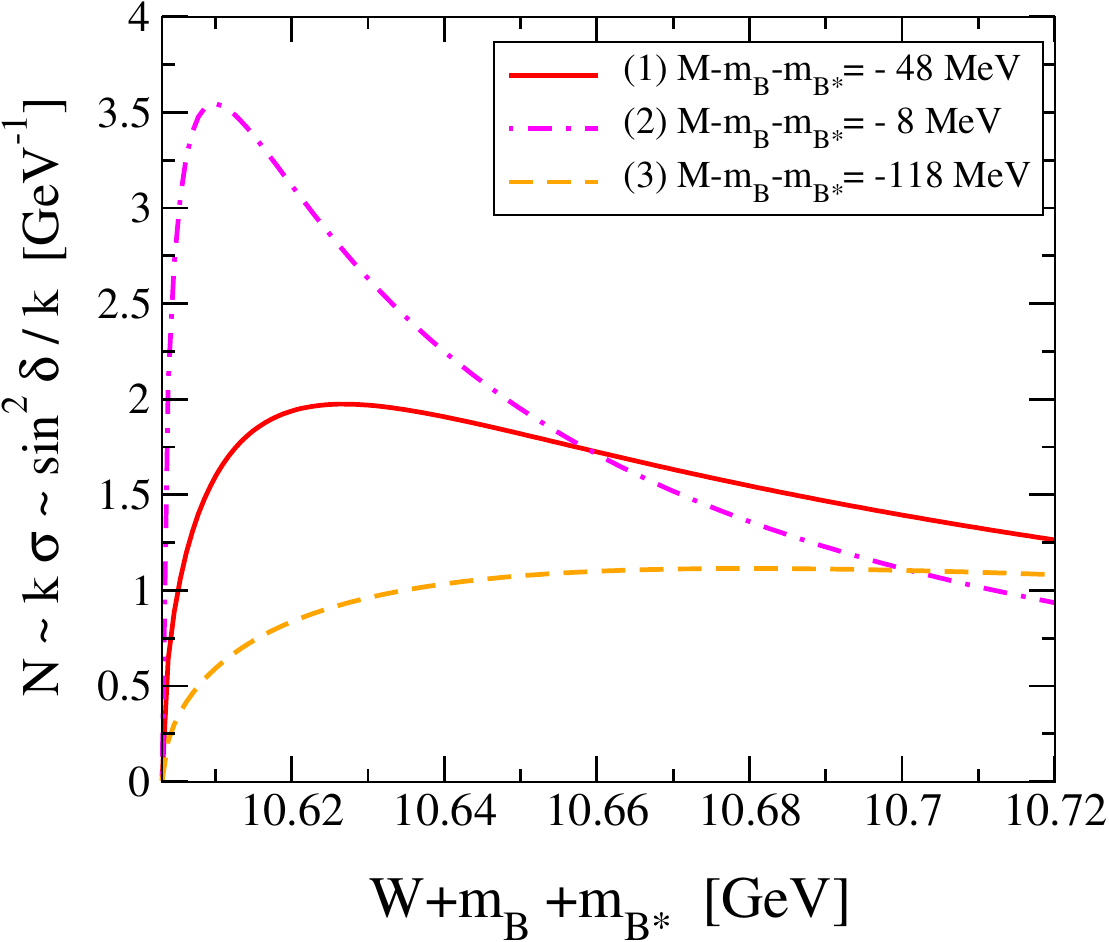}  
\caption{   The quantity $\sin^2\delta/k$ that is proportional to the $B\bar B^*$ rate $N_{B\bar B^*}\propto k \sigma_{B\bar B^*}$.  Three  choices of fits   consistent with  our lattice potential   are shown and the mass of the bound state is given for each case:   $(1)\ F=1.3,A=0.26,d=2.5$ (central value in Eq. \ref{mass}), $(2)\ F=2.0,A=0.15,d=2.8$ and $(3)\ F=1.0,A=1.0,d=1.0$.  } \label{fig:N}
\end{figure}

    This bound state  is responsible for a peak  in the $B\bar B^*$ rate $N_{B\bar B^*}\propto k \sigma\propto \sin^2\delta_{0}(W)/k$ above threshold if the bound state lies closely below threshold. This is illustrated in  Fig.  \ref{fig:N}, which  shows this rate  for three fits that are all consistent with our lattice potential:  the fit leading to the  central value   (\ref{mass})  and two choices of the fits that lead to a small or a large binding energy.   The bound state with a small binding energy $\simeq 8~$MeV leads to a peak in the $B\bar B^*$ rate  above threshold.    Its shape    resembles the $Z_b(10610)$ peak in the $B\bar B^*$ rate observed by Belle (Fig. 2 of \cite{Garmash:2015rfd}).  

 The significantly attractive $B\bar B^*$ potential (Fig. \ref{fig:V}) and the resulting  bound state in Fig.  \ref{fig:masses}   could be related to the existence of $Z_b$ in experiment. The   reliable relation between both will be possible only when simplifications employed here will be overcome in the future simulations.  The $Z_b(10610)$ was found as a virtual bound state  slightly below threshold  by the re-analysis of the experimental data  \cite{Wang:2018jlv} when the coupling to bottomonium light-meson channels was turned off \cite{Wang:2018jlv} (the position of the pole is only slightly shifted when this small coupling is taken into account).

  The exotic $Z_b$ resonances were observed only by Belle, so their confirmation by another experiment would be highly welcome. LHCb could try to search for it in inclusive final state $B\bar B^*$. 
  
   \vspace{0.1cm}

{\it Comparison with a previous lattice study:} Only one preliminary lattice study \cite{Peters:2016wjm} of this channel was reported up to now, considering heavier $m_\pi\!=\!480~$MeV and  twisted-mass fermions. It employed operators $O^{B\bar B^*}$ and $O^{\Upsilon\pi(0)}$, while $O^{\Upsilon\pi(p\not = 0)}$ and $O^{\Upsilon b_1(0)}$ were omitted. Two eigenstates are  interpreted as  $\Upsilon\pi(0)$ and  $B\bar B^*$.  The resulting potential $V(r)\!=\!E_{B\bar B^*}(r)-m_B-m_{B^*}$ (red line in Fig. 1 of \cite{Peters:2016wjm}) is also attractive and slightly  weaker than our potential for small $r/a\simeq 1,2$. This   difference  could be a consequence of a very different $m_\pi$. Their   potential and its parametrization  via $V(r)=-\tfrac{\alpha}{r}e^{-(r/d)^2}$   resulted in one bound state at   $M\!-\!m_{B}\!-\!m_{B^*}\!=\!-58 \pm 71~$MeV in \cite{Peters:2016wjm}. 
This is consistent with our binding energy shown in Fig.  \ref{fig:masses}.  Further lattice studies, including the comparison of the potentials at similar $m_\pi$, are highly awaited.
  
 \vspace{0.1cm}

{\it Conclusions:} We presented a lattice QCD study of  a channel with quark structure $\bar bb\bar du$, where Belle  observed two exotic $Z_b$ hadrons.    We find significantly attractive potential $V(r)$ between $B$ and $\bar B^*$ at small $r$ when the total spin of the heavy quarks is equal to one.  Dynamics of $B\bar B^*$ system within the extracted $V(r)$  leads to one bound state,  whose mass depends on the parametrization of $V$. Certain parametrizations  render  a bound state closely below threshold and  a narrow peak  in $B\bar B^*$ rate just above threshold, resembling  $Z_b$ in experiment.  

For quantitative comparison to experiment, future lattice studies need to explore how the  dynamics of $B\bar B^*$ is influenced by the coupling to $\Upsilon\pi$ channels, and by the component where the total spin of the heavy quarks is equal to zero. Derivation of the appropriate  analytic form for $V(r)$ would be very valuable. 

\vspace{0.3cm}

\textbf{Acknowledgments} 

\vspace{0.1cm}

We thank   G. Bali,   V. Baru, P. Bicudo,  N. Brambilla, E. Braaten, C. Hanhart, M. Karliner,  R. Mizuk, A. Peters and  M. Wagner for valuable discussions.    S.P. acknowledges support by Research Agency ARRS (research core funding No. P1-0035 and No. J1-8137) and DFG grant No. SFB/TRR 55. H.B. acknowledges support from the Scientific and Technological Research Council of Turkey (TUBITAK) BIDEB-2219 Postdoctoral Research Programme. 

 \vspace{1cm}
  
  {\it Note added:} In the previous version of this manuscript, the sub-matrix of the correlation matrix $C_{ij}$ was multiplied by a wrong overall constant (that  applied for  the sub-matrix related to $O_{3-6}$ that was averaged over source time slices). This affected the eigen-energies and overlaps. All figures (expect for Fig. 1) are replaced and are based on the correct correlation matrices, while majority of text remains unmodified. The main conclusion does not change - $Z_b$ is related to    attraction between $B$ and $\bar B^*$ at small distance. The  correct   potential is less attractive   and it results only in one bound state   that  is related to  $Z_b$ (the previous version quoted one virtual bound state slightly below threshold and an unexpected bound state far below threshold).

\newpage

\renewcommand{\thesection}{S\arabic{section}}   
\renewcommand{\thetable}{S\arabic{table}}   
\renewcommand{\thefigure}{S\arabic{figure}} 
\renewcommand{\theequation}{S\arabic{equation}} 

\setcounter{equation}{0}
\setcounter{figure}{0}
\setcounter{table}{0}

 \newpage
 
 \newpage

\centerline{\large \bf  SUPPLEMENTARY INFORMATION}

 
 \section{ S1: Symmetries and operators}

In this section we provide more details on the transformation properties   of the investigated system in Fig. 1(a) with  quantum numbers   in Eq. (2). Transformations  and quantum numbers are considered     on the example of operators $O_1$ and 
$O_4$ (3). The first line in $O_1$ separates Dirac indices of the heavy and light quarks, which simplifies specific transformations. 

The z-component of the  angular momentum $J_z^{l}=0$ is the eigenvalue related to the rotation of the light degrees of freedom around z-axes. The light-quark part of $O_1$ is $\bar q^b_B\Gamma_{BA}q_A^a\propto\bar q^b (1-\gamma_t)\gamma_5q^a$, which  has    angular momentum equal to zero indeed.   The light degrees of freedom in  $O_4$  are represented by  the pion with  momentum $ \vec p \propto e_z$ and a  straight gauge link path $U$ between 0 and r. Both have z-component of the angular momentum equal to zero.

The quantum number $\epsilon\!=\!-1$ is related to the reflection of the light-degrees of freedom over $yz$ plane, which is a product of rotation $R_{x,\pi}$ by $\pi$ around  x and inversion  $I$ with respect to the midpoint between 0 and r. The light-quark part $\bar q^b (1-\gamma_t)\gamma_5q^a$ of $O_1$ is invariant under rotations and has $P=-1$, therefore $\epsilon=-1$. The pion with momenta in z-direction within $O_4$ transforms as $\pi_{\vec p=\vec e_z}\stackrel{R_{x,\pi}}{\longrightarrow} \pi_{-\vec e_z}\stackrel{I}{\longrightarrow}-\pi_{\vec e_z}$, while the straight gauge link is invariant under this reflection, so $\epsilon=-1$. 

The Dirac structure for the heavy quark part  in all operators is $\bar b \gamma_z P_+ b$, which ensures $S^{h}=1$ and $S_z^{h}=0$. 

 The  $C\!\cdot\! P=-1$  is related to the product of the charge-conjugation and inversion with respect to the mid-point between 0 and r. Both  refer to the transformation of the light-degrees of freedom as well as the transformation of the static color sources\footnote{ If the color of the static source was not transformed under the charge-conjugation, the color-singlet $\bar b^a q^a \bar q^b b^b$  would transform under $C$-conjugation to $\bar b^a C\bar q^{aT} q^{bT}C b^b$, which is not gauge invariant. }. This   is most conveniently accomplished by the usual transformation rules $\psi \stackrel{C}{\to}C\bar \psi^T$ and $\psi \stackrel{P}{\to}\gamma_t \psi$  for both $\psi=q$ and $b$, where this operation does not affect the  heavy quark spin, while  $C=i\gamma_2\gamma_t$.  The operator $O_1$ has $C\!\cdot\! P=-1$ since 
 \begin{align}
 O_1&= \sum_{a,b}  \bar b^a(0) \tilde \Gamma b^b(r) ~~\bar q^b(r) \Gamma  q^a(0)  \\
 & \stackrel{C}{\longrightarrow}   \sum_{a,b} b^{aT}(0) C \tilde \Gamma C \bar b^{bT}(r) ~~ q^{bT}(r) C \Gamma C  \bar q^{aT}(0)\nonumber \\
 & =\sum_{a,b}\bar b^{b}(r) C \tilde \Gamma^T C   b^{a}(0) ~~ \bar q^{a} (0) C \Gamma^T C q^{b}(r) \nonumber\\
 \stackrel{P}{\longrightarrow} &\sum_{a,b}\bar b^{b}(0) \gamma_t C \tilde \Gamma^T C \gamma_t  b^{a}(r) ~~ \bar q^{a}(r) \gamma_t C \Gamma^T C \gamma_t q^{b}(0) =-O_1~,\nonumber
 \end{align}
where $P$ exchanges positions $0$ and $r$, $\gamma_t C \tilde \Gamma^T C \gamma_t=\tilde \Gamma$ for $\tilde \Gamma=\gamma_zP_+$,  $\gamma_t C \Gamma^T C \gamma_t=-\Gamma$ for $\Gamma=P_-\gamma_5$,   and dummy indices $a\leftrightarrow b$ can be exchanged in the last expression.  The linear combination $\pi_{\vec p=\vec e_z}+ \pi_{\vec p=-\vec e_z}\stackrel{C}{\longrightarrow}\pi_{\vec e_z}+ \pi_{-\vec e_z}\stackrel{P}{\longrightarrow}-\pi_{-\vec e_z}- \pi_{\vec e_z}$ ensures good $C\!\cdot\! P=-1$ for $O_4$.  

\vspace{0.1cm}

 The second line in operator $O_1$ (3) is obtained via the Fierz rearrangement
 \begin{equation}
 \tilde \Gamma_{CD} \Gamma_{BA}=\frac{1}{16}\sum_{\Gamma^1,\Gamma^2} \mathrm{Tr}[\Gamma^1 \tilde \Gamma \Gamma^2 \Gamma] \Gamma^1_{CA} \Gamma^2_{BD} 
 \end{equation} 
 and further simplifies since static heavy quarks appear in the combinations $P_+b$ and $\bar b P_-$.

\begin{figure*}[htb!]
\includegraphics[width=0.52\textwidth]{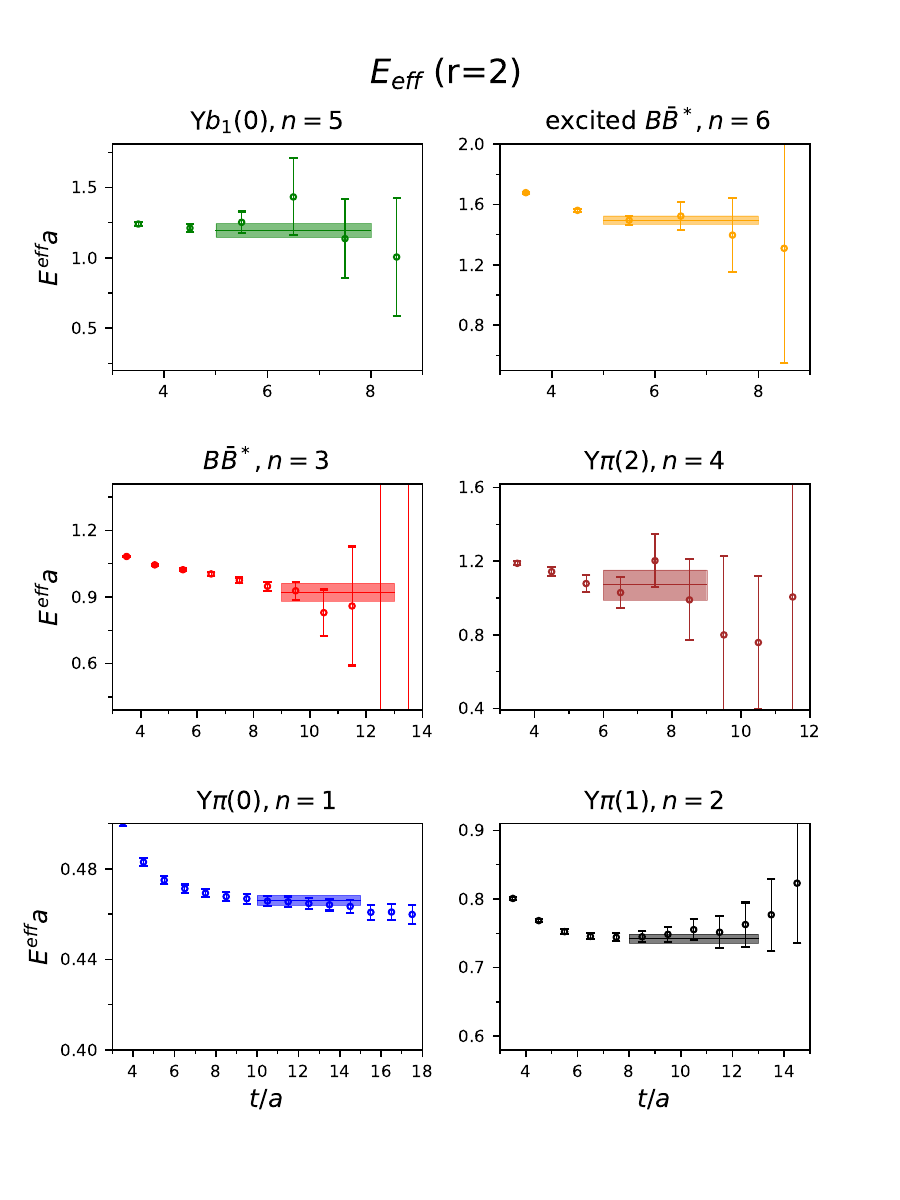}  $\ $
\includegraphics[width=0.46\textwidth]{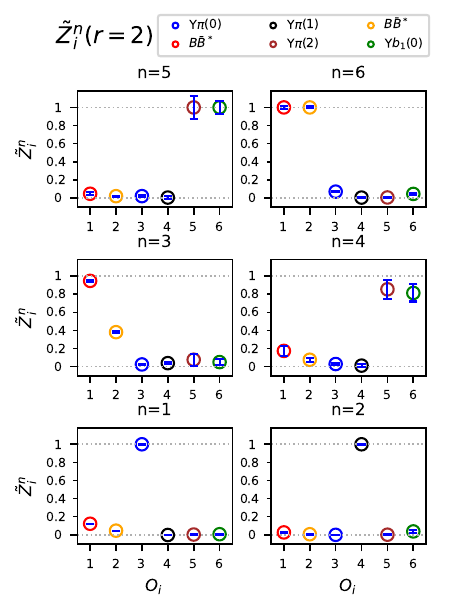}
\caption{  (a)  Effective energies $E_n^{eff}$ of the system in Fig. 1a for  separation $r/a=2$ and all eigenstates $n=1,..,6$. They render eigen-energies $E_n$ in the plateau region. (b) Normalized overlaps $\tilde Z_i^n\propto \langle O_i|n\rangle$ of each eigenstate $n$ on the left to six operators $O_{i=1,..,6}$. Absolute values of overlaps are shown for $r/a=2$. } \label{fig:Eeff}
\end{figure*}
 
\section{ S2: Effective energies and overlaps}

Effective energies $E_n^{eff}$ of the system in Fig. 1a are shown in Fig. \ref{fig:Eeff}(a) for  separation $r/a=2$ and all eigenstates $n=1,..,6$.  They are obtained from the correlation matrices  $C_{ij}(t)$ via variational approach $C(t)u_n(t)=\lambda_n(t)  C(t_0)u_n(t)$, where the effective energies are given by the eigenvalues  $E_n^{eff}(t)\equiv\ln[\lambda_n(t)/\lambda_n(t+1)]$. Reference time $t_0=[2,4]$ is used for various $r$ and agreement in this range of $t_0$ is verified. Effective energies render eigen-energies $E_n$ in the plateau region, indicated in the plots. 

The overlaps $\langle O_i|n\rangle$ of each eigenstate $n$ to employed operators $O_i$ (3) are shown in terms of the normalized overlaps $\tilde Z_i^n$ in Fig. \ref{fig:Eeff}(b). Here $\tilde Z_i^n\equiv \langle O_i|n\rangle/\max_m \langle O_i|m\rangle$ is normalized so that its maximal value for given $O_i$  across all eigenstates is equal to one.  

\begin{figure*}[htb!]
\includegraphics[width=0.47\textwidth]{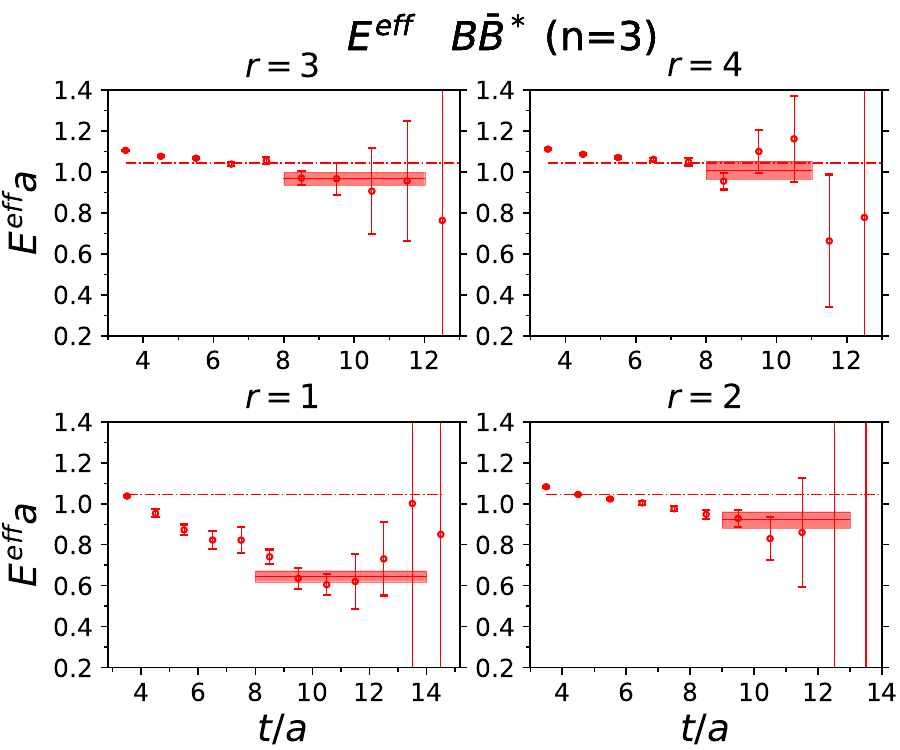} $\ $
\includegraphics[width=0.49\textwidth]{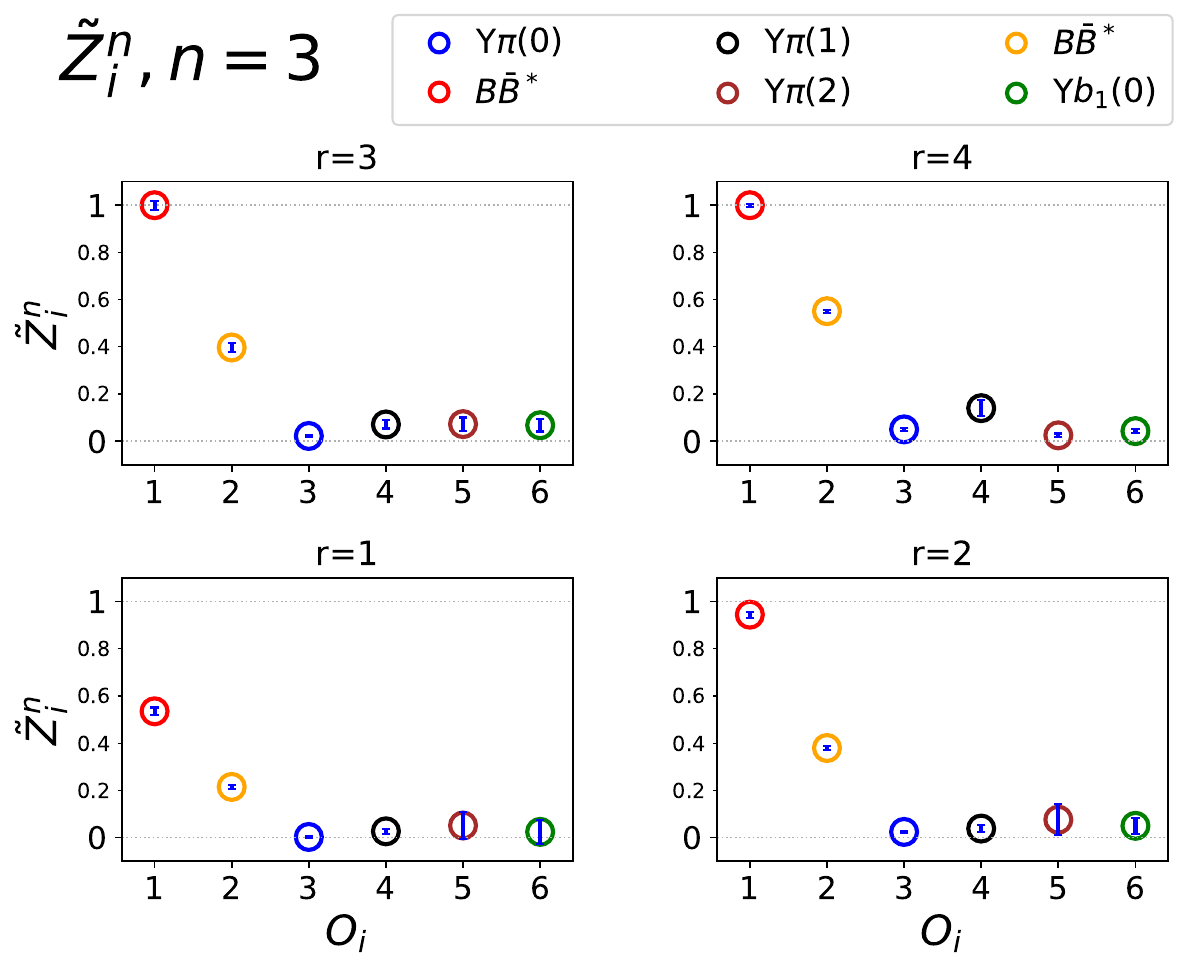} 
\caption{   Eigenstate dominated by $B\bar B^*$ (red circles in Fig. 2)  for separations $r/a=[1,4]$: (a) effective energies $E^{eff}(t)$ and (b)   overlaps $\tilde Z_i^n\propto \langle O_i|n\rangle$, where the absolute value of the overlap is shown.  } \label{fig:overlaps_log}
\end{figure*}

The effective energies (with fits) and overlaps of the eigenstate dominated by $B\bar B^*$ (red circles in Fig. 2) are presented  in Fig. \ref{fig:overlaps_log}.

The previous paragraphs apply to the case when all six operators listed in (3) are employed.  These operators interpolate all five two-hadrons states $B\bar B^*,~\Upsilon \pi(\vec p=0,1,2),~\Upsilon b_1(\vec p=0)$ that are expected in the relevant energy region below $m_B+m_{B^*}+0.2~$GeV.   If one added further operators, for example   replacing $\gamma_5$ by $\gamma_5\gamma_t$ within $O^{\Upsilon\pi}$ or employing    different smearing on the light quarks $q$, the resulting eigenvalues   would reach plateau earlier in time. However it is expected that the resulting eigen energies would be consistent with the ones obtained here; the exception could be the highest eigenstate indicated by yellow  in Fig. 2 that is  above  our region of interest and is not employed in the interpretation of result here.  

We investigated the eigen-energies and overlaps if some of the operators among (3) are omitted from the  correlation matrix.  Note that the results based on omitting any of $O^{\Upsilon\pi}$ or $O^{\Upsilon b_1}$ can not be regarded as  reliable, since they lie below or near eigenstate $B\bar B^*$ of interest.   We found, for example, that $\Upsilon\pi(2)$ eigenstate (denoted by brown in Fig. 2) disappears from the spectrum if operator $O^{\Upsilon\pi(2)}$  is omitted.  Similarly   $\Upsilon\pi(0)$ eigenstate seem to disappear from the spectrum at intermediate $t$ when $O^{\Upsilon\pi(0)}$  is omitted, but  some of the effective energies show falling behavior at larger $t$, indicating a coupling to the lightest eigenstate $\Upsilon\pi(0)$.  We note  that all the results in  this work are based on including all operators (3). 

\section{S3: Potential between $B$ and $\bar B^*$ from lattice  }

The lattice potential $V(r)$ between $B$ and $\bar B^*$ from Fig. 3a is tabulated in Table \ref{tab:V}. 

\begin{table}[h!]
\begin{tabular}{c|c}
      $r/a$    &  $V(r) a$   \\
      \hline 
      1 & $ -0.399 \pm 0.027$\\
     2 &  $-0.123 \pm 0.039$\\
   3 &   $-0.078 \pm 0.031$ \\
   4 & $-0.036 \pm 0.044$
      \end{tabular}
\caption{Potential $V(r)$ between $B$ and $\bar B^*$ extracted from our simulation and plotted in Fig. 3a. Potential $V(r)$ for separations $r/a>4$ is equal to zero within statistical and systematic errors. }\label{tab:V}
\end{table}

\section{S4:  Potential between $B$ and $\bar B^*$ at very small separation $r$}

Let us consider  the potential between $B$ and $\bar B^*$ analytically, where $b$ and $\bar b$ are separated by a very small distance  $r\ll r_B$, such that $r$ is  much smaller than average distance $r_B$ between $b$ and $\bar q$ in $B^{(*)}$ meson (i.e. average radius $r_B$ of a static $B^{(*)}$ meson).   We address the question whether this potential has a singular form $V_{1/r}(r)= \frac{K}{r}$ for $r\to 0$  and determine prefactor $K$, while we omit all sub-leading contributions that are finite at $r\to 0$.  Among all pairs of the four quarks $\bar b b \bar q q$, only the interaction between $b$ and $\bar b$ at very small  $r$ could give potential proportional to $1/r$. All other pairs are  at average  distance of the order of $O(r_B)$, which is  finite for $r\to 0$; these pairs do not lead to infinite potential for $r\to 0$ and we therefore omit their contribution to $V_{1/r}$. 

The task is therefore to determine potential between $b$ and $\bar b$ within  a pair of color-singlet  $B^{(*)}$ mesons  
\begin{equation}
\label{1}
|B \bar B^*\rangle=\tfrac{1}{\sqrt{3}} (\bar b q)  \tfrac{1}{\sqrt{3}}(\bar q b)  = \tfrac{1}{3} \sum_{a=1,3}\sum_{b=1,3}\bar b_a q_a  ~\bar q_b b_b~.
\end{equation}
The color structure with indices  $a$ and $b$  matches with the employed operators $O^{B\bar B^*}$, while other indices will not be relevant below.  In order to determine the potential,   $| B \bar B^*\rangle$ is expressed in terms of color singlets and octets  
\begin{equation}
|B \bar B^*\rangle=
\tfrac{1}{3}~\bigl\{ (\tfrac{1}{\sqrt{3}} \bar b b)(\tfrac{1}{\sqrt{3}} \bar q  q)+\sum_{A=1,..,8} \!\!\!( \tfrac{1}{\sqrt{2}} \bar b \lambda_A b)(\tfrac{1}{\sqrt{2}} \bar q  \lambda_A q) \bigr\}~.\nonumber
\end{equation} 
The $\bar bb$ singlet within $\langle B\bar B^*|B \bar B^*\rangle$ renders  the singlet  potential $V_0(r)$, all eight octets render octet potential $V_8(r)$
 \begin{align}
&\langle \tfrac{1}{\sqrt{3}} \bar b b  | \tfrac{1}{\sqrt{3}} \bar b b \rangle \to V_0(r)=-\frac{4}{3} \frac{\alpha_s}{r} +{\cal O}\bigl(\frac{\alpha_s^2}{r}\bigr),\\
&\langle \tfrac{1}{\sqrt{2}} \bar b\lambda_A b  | \tfrac{1}{\sqrt{2}} \bar b\lambda_A b \rangle\to V_8(r)=\frac{1}
{6} \frac{\alpha_s}{r} +{\cal O}\bigl(\frac{\alpha_s^2}{r}\bigr),\ A=1,..,8,\nonumber
\end{align}
while  $\langle \tfrac{1}{\sqrt{3}} \bar q q | \tfrac{1}{\sqrt{3}} \bar q  q \rangle\to 1$ and    $\langle \tfrac{1}{\sqrt{2}} \bar q \lambda_A q | \tfrac{1}{\sqrt{2}} \bar q \lambda_A q \rangle\to 1$ are properly normalized to one. 
The resulting $B\bar B^*$ potential at very small $r$ is therefore
\begin{align}
\label{V1/r}
V_{1/r}(r)&=\tfrac{1}{9} [V_0(r)+8 V_8(r)], \quad  V_{1/r}^{{\cal O}(\alpha_s)}(r)=0,\nonumber\\
  V_{1/r}(r)&=\frac{1}{9} \frac{4}{3} \frac{\alpha_s}{r}  \biggl(\frac{\alpha_s}{4\pi}\biggr)^2 ~\delta a_2 ~.
\end{align}
The singlet and octet contributions cancel in the case of one-gluon exchange, i.e at the order $O(\alpha_s)$. The lowest non-zero contribution  can be obtained from the perturbative calculation of both potentials in \cite{Kniehl:2004rk} and comes at $O(\alpha_s^3)$ with $\delta a_2=-189.2  $. Employing the value of $\alpha_s\simeq 0.31$ obtained from the fit of the singlet $\bar bb$ static potential in our simulation, we arrive at      $V_{1/r}(r)\simeq -0.0051/r$.

\begin{figure*} [htb]
\includegraphics[width=0.65\textwidth]{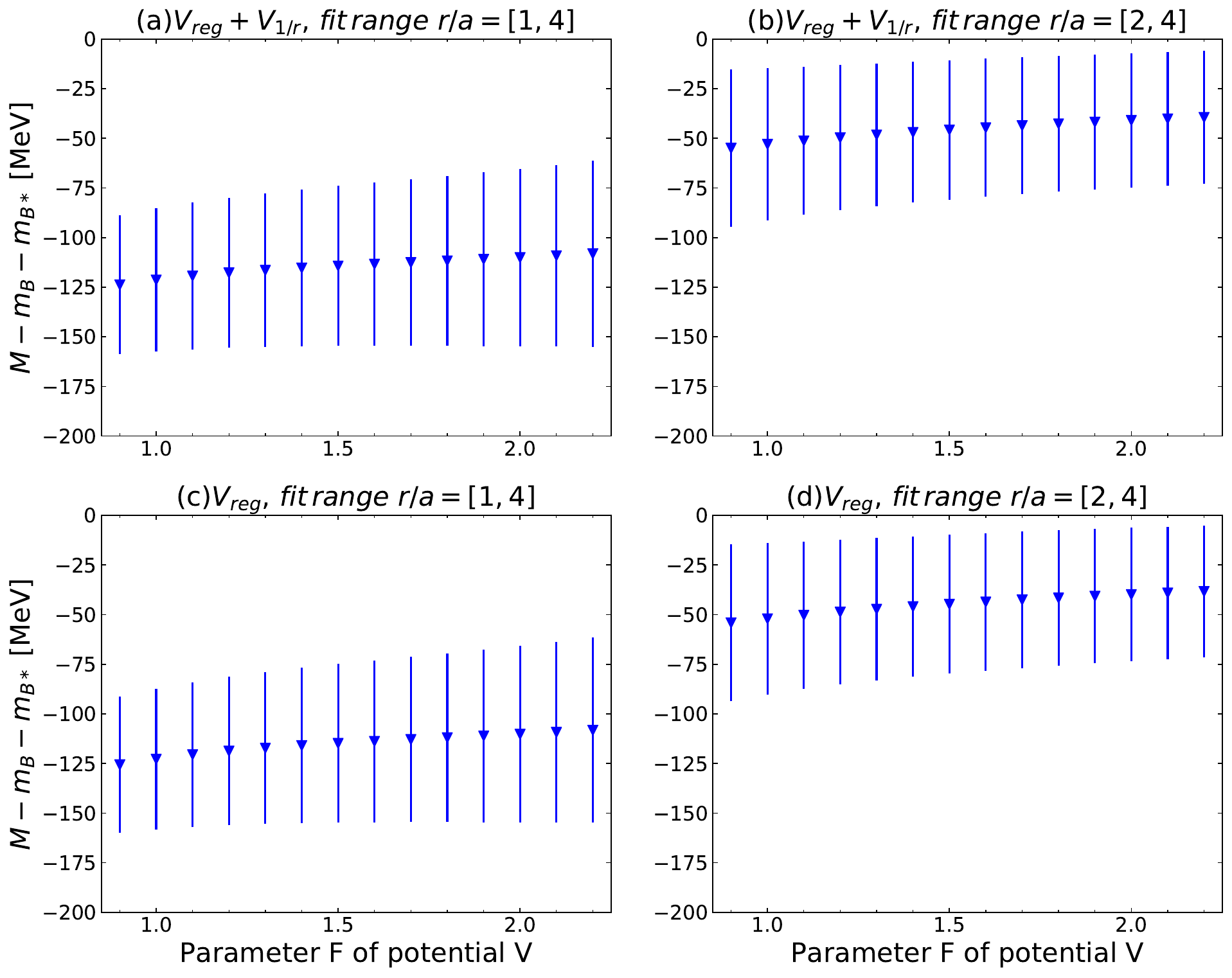} 
\caption{   Mass of the virtual bound state for various choices of the parameter $F$ in $V(r)$ (5). The plot compares results based on the fit of the lattice potentials in the ranges $r/a=[1,4]$ (left) and $r/a=[2,4]$ (right).  The results based on the potential $V_{reg.}+V_{1/r}$ (top) and $V_{reg.}$ (bottom) are also compared. } \label{fig:various_fits}
\end{figure*}
\section{S5: Mass of the bound state for various fits}

The mass of the bound state depends on the choices of the fit for the potential. 
The  mass   in Fig. 4 of the main article was based on the fits of the lattice potential in the ranges $r/a=[2,4]$, $r/a=[1,4]$ and form of potential $V(r)=V_{reg.}(r)+V_{1/r}(r)$ in Eq. (5). These masses are shown again in Fig.  \ref{fig:various_fits}(a,b) for completeness.  Two fitting ranges were employed since  the lattice potential at $r/a=1$ can be prone to the lattice discretization errors.   

The part of the potential $V_{1/r}(r)$ that is singular as $r\to 0$ was determined perturbatively (\ref{V1/r}) for very small separations $r$. It is equal to zero at the one-gluon exchange level and the lowest non-zero contribution comes at the order $O(\alpha_s^3)$.  The sensitivity of the results on including or excluding this part of the potential is explored in Fig.  \ref{fig:various_fits}. The masses based solely on the regular  potential $V_{reg.}$ in Figs.  \ref{fig:various_fits}(c,d) agree within errors with masses based on $V_{reg.}+V_{1/r}$ in Figs.  \ref{fig:various_fits}(a,b). This agreement is a consequence of the suppression in   $V_{1/r}$.

 \end{document}